\documentclass{amsart}
\usepackage{amsmath}

\newtheorem{lemma}{\bf Lemma}[section]
\newtheorem{theorem}{\bf Theorem}[section]
\newtheorem{corollary}{\bf Corollary}[section]
\newtheorem{proposition}{\bf Proposition}[section]
\newtheorem{example}{\bf Example}[section]

\newtheorem{remark}{\bf Remark}

\def\La{\Leftarrow}
\def\o{\overline}
\def\laa{\leftarrow}
\def\Ra{\Rightarrow}
\def\Pr{{\bf Proof.}\rm }
\def\fbx{\hfill${}^{\rule{2mm}{2mm}}$}
\begin{document}

\email{shtrakov@aix.swu.bg}
\address{Department of Computer Science \\
South-West University,  2700 Blagoevgrad, Bulgaria}
\title[ESSENTIAL ARITY GAP OF BOOLEAN
FUNCTIONS]
 {ESSENTIAL ARITY GAP OF BOOLEAN
FUNCTIONS}
\author[Sl. Shtrakov]{Slavcho Shtrakov}
\urladdr{http://home.swu.bg/shtrakov}
\date{}

\begin{abstract}

 In this paper we investigate the Boolean  functions with maximum essential arity gap.
  Additionally we propose a simpler prove
of an important theorem  proved by M. Couceiro and E. Lehtonen in
\cite{mig}. They used the Zhegalkin's polynomials as normal forms
for Boolean functions and describe the functions with essential
arity gap equals $2$. We use Full Conjunctive Normal Forms instead
of these polynomials which
 allow us to simplify the proofs and to obtain several
combinatorial results, concerning the Boolean functions with a given
arity gap. The Full Conjunctive Normal Forms  are sum of
conjunctions also, in which  all variables occur.

\end{abstract}
\maketitle

\section{Introduction}

Essential variables of functions are studied by several authors
\cite{bre,ch51,sal}.  In this paper we consider the problem for
 simplification of functions by identification of variables.
This problem is  discussed in the work of O. Lupanov, Yu. Breitbart,
A. Salomaa, M. Couceiro, E. Lehtonen,  etc. for Boolean functions
and by   K. Chimev for arbitrary discrete functions. Similar
problems for terms and universal algebra are studied by the author
and K. Denecke \cite{sh51}. Essential input variables for tree
automata are discussed  in \cite{s1}. The problems concerning
essential arity gap of functions are discussed in \cite{mig}. Here
we study and count the Boolean functions, which have maximum arity
gap. Note that if a function $f$ has greater essential arity gap
than the essential arity gap of another function $g$, then $f$ has a
simpler automata realization than $g$. This fact is of a great
importance  in theoretical and applied computer science and
modeling.

\section{Essential variables in Boolean functions}

 Let  $B=\{0,1\}$ be the set (ring) of
the residuums modulo $2$. An {\it $n$-ary Boolean function
(operation) } is a mapping $f: B^n\to B$ for some natural number
$n$, called {\it arity} of $f$. The set of  all such functions is
denoted by $P_2^n$.

A variable $x_i$ is called {\it essential} in $f$, or $f$ {\it
essentially depends} on $x_i$, if there exist values
$a_1,\ldots,a_n,b\in B$, such that
$$
   f(a_1,\ldots,a_{i-1},a_{i},a_{i+1},\ldots,a_n)\neq
   f(a_1,\ldots,a_{i-1},b,a_{i+1},\ldots,a_n).
$$

The set of essential variables in a function $f$ is denoted by
$Ess(f)$ and the number of essential variables in $f$ is denoted by
$ess(f)=|Ess(f)|$. The variables from $X=\{x_1,\ldots,x_n\}$ which
are not essential in
 $f\in P_2^n$ are called {\it fictive} and the set of fictive
 variables in $f$ is denoted by $Fic(f)$.

Let $x_i$ and  $x_j$ be essential variables in $f$. We say that the
function $g$ is obtained from $f\in P_{2}^{n}$ by {\it
identification of a variable $x_i$ with $x_j$}, if
$$
   g(x_1,\ldots,x_n)=f(x_1,\ldots,x_{i-1},x_j,x_{i+1},\ldots,x_n)=f(x_i=x_j).
$$
Briefly, when  $g$ is obtained from $f, $ by identification of the
variable $x_i$ with $x_j$, we will write $g=f_{i\laa j}$ and $g$ is
called {\it the identification minor of $f$}. The set of all
identification minors of $f$ will be denoted by $Min(f)$.

For completeness of our consideration we alow to obtain
identification minors when $x_i$ or $x_j$ are not essential in $f$,
also. Thus if $x_i$ does not occur in $f$, then we define $f_{i\laa
j}:=f$.

Clearly, $ess(f_{i\laa j})\leq ess(f)$, because   $x_i\notin
Ess(f_{i\laa j})$, even though it may be essential in $f$.

For a function $f\in P_2^n$ the {\it essential arity gap} (shortly
{\it arity gap}) of $f$ is defined as follows
$$gap(f):=ess(f)-\max_{g\in Min(f)}ess(g).$$

It is not difficult to see that the functions with "huge" gap, are
more simple for realization by switching circuits and functional
schemas in theoretical and applied computer science.

Let us denote by $G_p^m$ the set of all functions in $P_2^n$ which
essentially  depend on $m$ variables and have gap equals to $p$ i.e.
$G_p^m=\{f\in P_2^n\ |\ ess(f)=m\ \&\ gap(f)=p\}$, with $m\leq n$.

An upper bound of $gap(f)$ for Boolean functions is found by
K.Chimev, A. Salomaa and  O. Lupanov  \cite{ch51,sal,lup}. It is
shown that  $gap(f)\leq 2$, when  $f\in P_2^n,\ n\geq 2$.

This result is generalized for arbitrary finite valued functions  in
\cite{mig}. It is proved that $gap(f)\leq k$ for all $f\in P_k^n$,
 $n\geq k$.

Let $m\in N$, $0\leq m\leq 2^n-1$ be an integer. It is well known
that for every $n\in N,\ $ there is an unique finite sequence
$(\alpha_1,\ldots,\alpha_n)\in B^n$ such that
\begin{equation}\label{eq3}
  m=\alpha_1 2^{n-1}+ \alpha_2 2^{n-2}+\ldots+\alpha_n.
\end{equation}
The equation (\ref{eq3}) is known as the presentation of $m$ in
binary  positional numerical system. One briefly  writes   $
m=\o{\alpha_1\alpha_2\ldots\alpha_n}$ instead (\ref{eq3}).

For a variable $x$ and $\alpha\in B$, we define the following
important function
$$
  x^\alpha=\left\{\begin{array}{ccc}
             1 \  &\  if \  &\  x=\alpha \\
             0 & if & x\neq\alpha.
           \end{array}
           \right.
$$
This function is used in many investigations, concerning the
applications of discrete functions in computer science \cite{ch51}.

There are many normal forms for representation of functions from
$P_2^n$.  In this paper we will use the {\it Full Conjunctive Normal
Form (FCNF)} for studying the essential arity gap of functions. This
normal form is based on the table representation of Boolean
functions.

The next two theorems are in the basis of the Theory of Boolean
functions, and they are well known.
\begin{theorem}\label{t1}
Each function $f\in P_2^n$ can be uniquely represented in FCNF as
follows
\begin{equation}\label{eq5}
    f(x_1,\ldots,x_n)=a_0.x_1^{0}\ldots x_n^{0}\oplus\ldots
     a_{m}.x_1^{\alpha_1}\ldots
     x_n^{\alpha_n}\oplus\ldots a_{2^n-1}.x_1^{1}\ldots
     x_n^{1}
\end{equation}
where  $m={\o{\alpha_1\ldots\alpha_n}}$, $a_{m}\in B$ and
$"\oplus"$, and  $"."$ are the operations addition and
multiplication modulo $2$ in the ring $B$.
\end{theorem}

\begin{theorem}\label{t2}
A variable $x_i$ is fictive in the function  $f\in P_2^n$,  if and
only if
$$
    f(x_1,\ldots,x_n)=$$ $$=
    x_i^{0}.f_1(x_1,\ldots,x_{i-1},x_{i+1},\ldots,x_n)\ \oplus\ x_i^{1}.f_2(x_1,\ldots,x_{i-1},x_{i+1},\ldots,x_n),
$$
with $f_1=f_2$ and $x_i\notin Ess(f_j)$, where $f_j\in P_2^{n-1}$,
for $j=1,2$.
\end{theorem}

The next lemmas characterize the relation between the identification
minors of Boolean functions.

\begin{lemma}\label{l3}
Let $f, g\in P_2^n$ be two Boolean functions represented by their
FCNF as follows
$$f=\bigoplus_{m=0}^{2^{n-1}-1}a_m.x_1^{\alpha_1}\ldots
 x_n^{\alpha_n}\ \mbox{ and
}\  g=\bigoplus_{m=0}^{2^{n-1}-1}b_m.x_1^{\alpha_1}\ldots
 x_n^{\alpha_n},$$ where $m=\o{\alpha_1\ldots\alpha_n}$.
If   $f_{i\laa j}=g_{i\laa j}$ and
 $\alpha_i=\alpha_j$ for some $i,\ j$\ with  $1\leq j<i\leq
n$, then $a_m=b_m.$
\end{lemma}
\Pr\  Without loss of generality we will prove the lemma, for $i=2$
and $j=1$. Since $f_{2\laa 1}=g_{2\laa 1}$ we have
$$f(x_1,x_1,x_3,\ldots, x_n)=g(x_1,x_1,x_3,\ldots, x_n).$$ Hence
$$a_m=f(\alpha_1,\alpha_1,\alpha_3,\ldots,
\alpha_n)=g(\alpha_1,\alpha_1,\alpha_3,\ldots, \alpha_n)=b_m.$$ \fbx

\begin{lemma}\label{l1}
Let $f,g\in P_2^n,\ $ be two  functions, depending essentially on
$n,\ n\geq 3$ variables. If   $f_{i\laa j}=g_{i\laa j}$ for all
$i,j,\ 1\leq j < i\leq n$, then $f=g$.
\end{lemma}
\Pr\ Let $f$ and $g$ be functions represented by  their FCNF as in
Lemma \ref{l3}. Let
$m=\alpha_1.2^{n-1}+\alpha_2.2^{n-2}+\ldots+\alpha_n$ be an
arbitrary integer from $\{0,1,\ldots,2^n-1\}$.
 Since $n\geq 3$  there exist two natural numbers $i,\ j$ with $1\leq j<i\leq
 n$ and $\alpha_i=\alpha_j$. From Lemma \ref{l3} we obtain
  $$a_m=f(\alpha_1,\alpha_2,\ldots,
\alpha_n)=g(\alpha_1,\alpha_2,\ldots, \alpha_n)=b_m.$$ Consequently,
we have  $f=g$.
\fbx
\begin{example}\label{e1}
Let us consider the Boolean functions $f=x_1^0x_2^0\oplus
x_1^1x_2^0$ and $g=x_1^0x_2^0\oplus x_1^0x_2^1$. It is easy to see
that for all $i,j,\ 1\leq j < i\leq n$ we have $f_{i\laa j}=g_{i\laa
j}=x_1^0$, but  $f\neq g$. This example shows that $n\geq 3$ is an
essential  condition in Lemma \ref{l1}.
\end{example}

\section{Essential Arity Gap of Boolean
Functions}

For Boolean functions  $\neg (x)$ denotes the unary operation
negation, i.e.
$$\neg{x}=x^0=\left\{\begin{array}{ccc}
             1 \  &\  if \  &\  x=0 \\
             0 & if & x\neq 0.
           \end{array}
           \right.$$

\begin{proposition}\label{p1}
For each Boolean function $f$ the following sentences are  held:

$(i)$\ $gap(f(x_1,\ldots,x_n))= gap(f(\neg x_1,\ldots,\neg x_n))$;

$(ii)$\ $gap(f(x_1,\ldots,x_n))= gap(\neg({f( x_1,\ldots, x_n)}))$;

$(iii)$\ $gap(f(x_1,\ldots,x_n))= gap({f( x_{\pi(1)},\ldots,
x_{\pi(n)})})$, where $\pi: \{1,\ldots,n\}\to \{1,\ldots,n\}$ is a
permutation of the set $\{1,\ldots,n\}$;

$(iv)$\  $ess(f_{i\laa j})= ess(f_{j\laa i})$ for all $i,j,\ 1\leq
j< i\leq n$.
\end{proposition}

Note that the last two assertions $(iii)$ and $(iv)$ are valid in
more general case of $k$-valued functions.

 For any natural number $n, n\geq 2$ we define the following two sets:
$$Od_2^n:=\{\alpha_1\alpha_2\ldots\alpha_n\in \{0,1\}^n\
|\ \alpha_1\oplus\alpha_2\oplus\ldots\oplus\alpha_n=1\}$$ and
$$Ev_2^n:=\{\alpha_1\alpha_2\ldots\alpha_n\in \{0,1\}^n\
|\ \alpha_1\oplus\alpha_2\oplus\ldots\oplus\alpha_n=0\}.$$ Clearly,
$\alpha_1\alpha_2\ldots\alpha_n\in Od_2^n$ if and only if the number
of 1's in $\alpha_1\alpha_2\ldots\alpha_n$ is odd, and
$\alpha_1\alpha_2\ldots\alpha_n\in Ev_2^n$ when this number is even.

\begin{proposition}~\label{p2}
For any $n$,  $n\geq 4$, if
$$f= \bigoplus_{\alpha_1\ldots\alpha_n\in
Od_2^n}x_1^{\alpha_1}\ldots
 x_n^{\alpha_n}\quad \mbox{ or }\quad
 f= \bigoplus_{\alpha_1\ldots\alpha_n\in Ev_2^n}x_1^{\alpha_1}\ldots x_n^{\alpha_n},$$
 then $f\in G_2^n$.
\end{proposition}
\Pr\ Without loss of generality let us  assume that  $f=
\bigoplus_{\alpha_1\ldots\alpha_n\in Od_2^n}x_1^{\alpha_1}\ldots
 x_n^{\alpha_n}$. We have to show that $ess(f_{i\laa j})\leq n-2$
 for all $i,j,\ 1\leq
j < i\leq n$.  Without loss of generality, again we will assume
$i=2$ and $j=1$. Then we have
$$f_{2\laa 1}=\bigoplus_{\alpha_1,\alpha_3,\ldots\alpha_n\in Od_2^{n-1}}
x_1^{\alpha_1}x_3^{\alpha_3}\ldots
 x_n^{\alpha_n}=$$ $$=x_1^0.\big[\bigoplus_{\alpha_3,\ldots\alpha_n\in Od_2^{n-2}}x_3^{\alpha_3}\ldots
 x_n^{\alpha_n}\big]\ \oplus\ x_1^1.\big[\bigoplus_{\alpha_3,\ldots\alpha_n\in Od_2^{n-2}}x_3^{\alpha_3}\ldots
 x_n^{\alpha_n}\big]=$$ $$=\bigoplus_{\alpha_3,\ldots\alpha_n\in Od_2^{n-2}}x_3^{\alpha_3}\ldots
 x_n^{\alpha_n}.$$
 The result is the same,  when $\alpha_1\ldots\alpha_n\in Ev_2^n$.
 \fbx

 We are going to describe the
set $G_2^n$ for $n=2,3,4$. The results for $n=4$ can be easily
extended in the more general case $n\geq 4$.
\begin{theorem}\label{t3}
Let $f\in P_2^2$. Then   $f\in G_2^2$ if and only if $$
f=a_0.(x_1^0x_2^0\ \oplus\ x_1^1x_2^1)\oplus a_1.x_1^0x_2^1\oplus
a_2.x_1^1x_2^0,\ \mbox{  with}\ a_1\neq a_0\ \mbox{ or }\ a_2\neq
a_0.$$
\end{theorem}
\Pr\ Let $f=a_0.x_1^0x_2^0\oplus a_1.x_1^0x_2^1\oplus
a_2.x_1^1x_2^0\oplus a_3.x_1^1x_2^1$. The variables $x_1$ and $x_2$
are essential in $f$ if and only if $(a_0,a_1)\neq (a_2,a_3)$ and
$(a_0,a_2)\neq (a_1,a_3)$. Consider the identification minor
$h:=f_{2\laa 1}=a_0.x_1^0\oplus a_3.x_1^1$ of $f$. We need
$ess(h)=0$ and from Theorem \ref{t2} it follows $a_0=a_3.$ If we
suppose that $a_1=a_2=a_0$, then $f(x_1,x_2)=a_0$ which contradicts
$ess(f)=2$. \fbx
\begin{corollary}\label{c3.1} There are 6 functions in $G_2^2$, i.e.
$|G_2^2|=6.$
\end{corollary}
\Pr\ Let $a_0\in\{0,1\}$. For $a_1$ and $a_2$ there are 3 possible
choices, which satisfy Theorem \ref{t3}. The both cases
$a_1=a_2=a_0=0$ and $a_1=a_2=a_0=1$ are impossible because then
$ess(f)<2$, since Theorem \ref{t2}. \fbx

\begin{corollary}\label{c3.3}
If $f=a_0.x_1^0x_2^0\oplus a_1.x_1^0x_2^1\oplus a_2.x_1^1x_2^0\oplus
a_3.x_1^1x_2^1\in P_2^2$ then $ess(f_{2\laa 1})=0$ if and only if
$a_0=a_3$.
\end{corollary}
The next step is to describe the functions which essentially depend
on 3 variables and have essential arity gap equal to 2.
\begin{theorem}\label{t4} Let
$f$ be a Boolean function of three variables. Then
 $f\in G_2^3$ if and only if it can be represented in one of the
 following special forms:
\begin{eqnarray}~~~\label{eq23}
 f=x_3^\alpha(x_1^0x_2^1\ \oplus\ x_1^1x_2^0)\
\oplus\ x_1^\beta x_2^{\beta},
\end{eqnarray}
or
\begin{equation}~~~\label{eq24}
f=x_3^\alpha(x_1^0x_2^0\ \oplus\ x_1^1x_2^1)\ \oplus\
x_3^{\neg(\alpha)}(x_1^0x_2^1\ \oplus\ x_1^1x_2^0),
\end{equation}

where $\alpha, \beta\in \{0,1\}$.
\end{theorem}
 \Pr\
Note that the presentation of $f$ in (\ref{eq24}) is symmetric with
respect to the variables, but in (\ref{eq23}) $f$ is not symmetric
with respect to the variables $x_1$ and $x_3$. So, the theorem
asserts that  $f\in G_2^3$ if and only if $f$ can be represented in
one of the forms
 (\ref{eq23}) or (\ref{eq24}), after a suitable permutation of the variables.

"$\La$" Clearly, $x_1,x_2$ and $x_3$ are essential variables in the
functions of the right sides of  (\ref{eq23}) and (\ref{eq24}). To
see that $f\in G_2^3$ it is enough to do immediate checking.   Thus
for the function $f$ in (\ref{eq23}) we have  $f_{2\laa
1}=x_1^\beta$,  $$f_{3\laa 1}=\left\{\begin{array}{ccc}
             x_1^\beta \  &\  if \  &\  \beta=\alpha \\
            x_2^\beta & if & \beta\neq\alpha
           \end{array}\right. \ \ \mbox{and}\ \ f_{3\laa 2}=\left\{\begin{array}{ccc}
             x_2^\beta \  &\  if \  &\  \beta=\alpha \\
            x_1^\beta & if & \beta\neq\alpha.
           \end{array}\right.$$
The functions $f$ as in (\ref{eq24}) are  in $G_2^3$ because
$x_i,x_j\notin Ess(f_{i\laa j})$ for all $i,j,\ 1\leq j < i\leq 3$.

"$\Ra$" Assume that $f\in G_2^3$. Let the FCNF of  $f$ is written as
follows:
\begin{eqnarray*}
f=x_3^0(a_0.x_1^0x_2^0\ \oplus\ a_1.x_1^0x_2^1\ \oplus\
a_2.x_1^1x_2^0\ \oplus\  a_3.x_1^1x_2^1)\ \oplus\\ \oplus
x_3^1(a_4.x_1^0x_2^0\ \oplus a_5.x_1^0x_2^1\ \oplus a_6.x_1^1x_2^0\
\oplus a_7.x_1^1x_2^1)=\\ =x_3^0.g(x_1,x_2)\ \oplus\
x_3^1.h(x_1,x_2).
\end{eqnarray*}

{\bf A.} Suppose that $x_1\in Ess(g_{2\laa 1})$ or $x_1\in
Ess(h_{2\laa 1})$. Then $x_1\in Ess(f_{2\laa 1})$ because $f_{2\laa
1}(x_3=0)=g_{2\laa 1}$ and $f_{2\laa 1}(x_3=1)=h_{2\laa 1}$. Hence
$f\in G_2^3$ implies $x_3\notin Ess(f_{2\laa 1})$ i.e $g_{2\laa
1}=h_{2\laa 1}$. Consequently, $a_0=a_4$ and  $a_3=a_7$. Then we
obtain
$$u=f_{3\laa 1}=a_0.x_1^0x_2^0\ \oplus\ a_1.x_1^0x_2^1\ \oplus\
a_6.x_1^1x_2^0\ \oplus\  a_7.x_1^1x_2^1,$$ and
$$v=f_{3\laa 2}=a_0.x_1^0x_2^0\ \oplus\ a_2.x_1^1x_2^0\ \oplus\
a_5.x_1^0x_2^1\ \oplus\  a_7.x_1^1x_2^1.$$  There are the following
cases:

{\bf A.a.} $x_1\notin Ess(u)$. Hence  $a_0=a_6$ and $a_1=a_7$.

{\bf A.a.1.} If we suppose that $x_1\notin Ess(v)$, then $a_0=a_2$
and $a_5=a_7$ implies (according Theorem \ref{t2}) that
$x_1,x_3\notin Ess(f)$ and $f\notin G_2^3$.

{\bf A.a.2.} If $x_2\notin Ess(v)$, then $a_0=a_5$ and $a_2=a_7$.
Note that if $a_0=a_7$, then $f$ has to be a constant. Hence
$a_7=\neg(a_0)$. Then we obtain
$$
f=a_0.\big[x_1^0x_2^0x_3^0\ \oplus\ x_1^0x_2^0x_3^1\ \oplus\
x_1^0x_2^1x_3^1\ \oplus\  x_1^1x_2^0x_3^1\big]\ \oplus$$ $$ \oplus\
\neg(a_0).\big[x_1^0x_2^1x_3^0\ \oplus x_1^1x_2^0x_3^0\ \oplus
x_1^1x_2^1x_3^0\ \oplus x_1^1x_2^1x_3^1\big]=$$
$$=a_0\big[x_3^1(x_1^0x_2^1\ \oplus\
x_1^1x_2^0)\ \oplus\ x_1^0x_2^0\big]\ \oplus\
\neg(a_0)\big[x_3^0(x_1^0x_2^1\ \oplus\ x_1^1x_2^0)\ \oplus\ x_1^1
x_2^1\big]\in G_2^3.$$ Clearly, $ f$ is presented as in
(\ref{eq23}).

{\bf A.b.} $x_2\notin Ess(u)$. Hence  $a_0=a_1$ and $a_6=a_7$.

 {\bf A.b.1.} If
we suppose that $x_2\notin Ess(v)$, then $a_0=a_5$ and $a_2=a_7$
implies (according Theorem \ref{t2}) that $x_2,x_3\notin Ess(f)$ and
$f\notin G_2^3$.

{\bf A.b.2.} If $x_1\notin Ess(v)$, then $a_0=a_2$ and $a_5=a_7$.
Again, if $a_0=a_7$, then $f$ has to be a constant. Hence
$a_7=\neg(a_0)$. Then we obtain
$$
f=a_0.\big[x_1^0x_2^0x_3^0\ \oplus\ x_1^0x_2^0x_3^1\ \oplus\
x_1^0x_2^1x_3^0\ \oplus\  x_1^1x_2^0x_3^0\big]\ \oplus$$ $$ \oplus\
\neg(a_0).\big[x_1^0x_2^1x_3^1\ \oplus x_1^1x_2^0x_3^1\ \oplus
x_1^1x_2^1x_3^0\ \oplus x_1^1x_2^1x_3^1\big]=$$
$$=a_0\big[x_3^0(x_1^0x_2^1\ \oplus\ x_1^1x_2^0)\ \oplus\
x_1^0 x_2^0\big]\ \oplus\ \neg(a_0)\big[x_3^1(x_1^0x_2^1\ \oplus\
x_1^1x_2^0)\ \oplus\ x_1^1 x_2^1\big]\in G_2^3.$$ Clearly, $ f$ is
presented as in  (\ref{eq23}).

 {\bf B.} Let us suppose that $x_1\notin Ess(g_{2\laa 1})$ and  $x_1\notin
Ess(h_{2\laa 1})$. Then we have $g\in G_2^2$ and $h\in G_2^2$.
 From  Theorem
\ref{t3} it follows that
$$g(x_1,x_2)=a_0.(x_1^0x_2^0\ \oplus\ x_1^1x_2^1)\ \oplus\ a_1.x_1^0x_2^1\ \oplus\
a_2.x_1^1x_2^0,$$ and
$$h(x_1,x_2)=a_4.(x_1^0x_2^0\ \oplus\ x_1^1x_2^1)\ \oplus\ a_5.x_1^0x_2^1\ \oplus\
a_6.x_1^1x_2^0.$$

 Then we obtain
$$u=f_{3\laa 1}=a_0.x_1^0x_2^0\ \oplus\ a_1.x_1^0x_2^1\ \oplus\
a_6.x_1^1x_2^0\ \oplus\  a_4.x_1^1x_2^1,$$ and
$$v=f_{3\laa 2}=a_0.x_1^0x_2^0\ \oplus\ a_2.x_1^1x_2^0\ \oplus\
a_5.x_1^0x_2^1\ \oplus\  a_4.x_1^1x_2^1.$$

{\bf B.a.} $x_1\notin Ess(u)$. Hence  $a_0=a_6$ and $a_1=a_4$.

{\bf B.a.1.}
 If $x_1\notin Ess(v)$,
then $a_0=a_2$ and $a_4=a_5$. Note that if $a_0=a_4$, then $f$ has
to be a constant. Hence $a_4=\neg(a_0)$. Then we obtain
$$
f=a_0.\big[x_1^0x_2^0x_3^0\ \oplus\ x_1^1x_2^0x_3^0\ \oplus\
x_1^1x_2^0x_3^1\ \oplus\  x_1^1x_2^1x_3^0\big]\ \oplus$$ $$ \oplus\
\neg(a_0).\big[x_1^0x_2^0x_3^1\ \oplus x_1^0x_2^1x_3^0\ \oplus
x_1^0x_2^1x_3^1\ \oplus x_1^1x_2^1x_3^1\big]=$$
$$=a_0\big[x_1^1(x_2^0x_3^1\ \oplus\
x_2^1x_3^0)\ \oplus\ x_2^0x_3^0\big]\ \oplus\
\neg(a_0)\big[x_1^0(x_2^0x_3^1\ \oplus\ x_2^1x_3^0)\ \oplus\ x_2^1
x_3^1\big]\in G_2^3.$$ Clearly, $ f$ is presented as in
(\ref{eq23}).

{\bf B.a.2.}
 If $x_2\notin Ess(v)$, then $a_0=a_5$
and $a_2=a_4$.  Again, if $a_0=a_4$, then $f$ has to be a constant.
Hence $a_4=\neg(a_0)$. Then we obtain
$$
f=a_0.\big[x_1^0x_2^0x_3^0\ \oplus\ x_1^0x_2^1x_3^1\ \oplus\
x_1^1x_2^0x_3^1\ \oplus\  x_1^1x_2^1x_3^0\big]\ \oplus$$ $$ \oplus\
\neg(a_0).\big[x_1^0x_2^0x_3^1\ \oplus x_1^0x_2^1x_3^0\ \oplus
x_1^1x_2^0x_3^0\ \oplus x_1^1x_2^1x_3^1\big]=$$
$$=a_0\big[x_3^0(x_1^0x_2^0\ \oplus\ x_1^1x_2^1)\ \oplus\
x_3^1(x_1^1x_2^0\ \oplus\ x_1^0x_2^1)\big]\ \oplus\ $$ $$\oplus\
\neg(a_0)\big[x_3^1(x_1^0x_2^0\ \oplus\ x_1^1x_2^1)\ \oplus\
x_3^0(x_1^1x_2^0\ \oplus\ x_1^0x_2^1)\big]\in G_2^3.$$ Clearly, $ f$
is presented as in (\ref{eq24}).

{\bf B.b.} $x_2\notin Ess(u)$. Hence  $a_0=a_1$ and $a_6=a_4$.

{\bf B.b.1.} If we suppose that $x_1\notin Ess(v)$, then $a_0=a_2$
and $a_4=a_5$ implies (according Theorem \ref{t2}) that
$x_1,x_2\notin Ess(f)$ and $f\notin G_2^3$.

 {\bf B.b.2.} If
$x_2\notin Ess(v)$, then $a_0=a_5$ and $a_2=a_4$.  Again, if
$a_0=a_4$, then $f$ has to be a constant. Hence $a_4=\neg(a_0)$.
Then we obtain
$$
f=a_0.\big[x_1^0x_2^0x_3^0\ \oplus\ x_1^0x_2^1x_3^0\ \oplus\
x_1^0x_2^1x_3^1\ \oplus\  x_1^1x_2^1x_3^0\big]\ \oplus$$ $$ \oplus\
\neg(a_0).\big[x_1^0x_2^0x_3^1\ \oplus x_1^1x_2^0x_3^0\ \oplus
x_1^1x_2^0x_3^1\ \oplus x_1^1x_2^1x_3^1\big]=$$
$$=a_0\big[x_2^1(x_1^0x_3^1\ \oplus\
x_1^1x_3^0)\ \oplus\ x_1^0x_3^0\big]\ \oplus\
\neg(a_0)\big[x_2^0(x_1^0x_3^1\ \oplus\ x_1^1x_3^0)\ \oplus\ x_1^1
x_3^1\big]\in G_2^3.$$ Clearly, $ f$ is presented as in
(\ref{eq23}).
 \fbx

\begin{corollary}\label{c4.1}
Let $f\in P_2^3$. Then $ess(f_{i\laa j})\leq 1$ for all $i,j,\ 1\leq
j < i\leq 3$ if and only if
\begin{eqnarray*}
f=x_3^\alpha(x_1^0x_2^0\ \oplus\ x_1^1x_2^1)\ \oplus\
a_1.x_1^0x_2^1x_3^0\ \oplus\  a_2.x_1^1x_2^0x_3^0\ \oplus\\ \oplus\
\neg({a_2}).x_1^0x_2^1x_3^1\ \oplus\  \neg({a_1}).x_1^1x_2^0x_3^1,
\end{eqnarray*}
where $\alpha, a_1, a_2\in\{0,1\}$.
\end{corollary}
\Pr\ This Corollary summarizes all cases considered in Theorem
\ref{t4}. For instance if $\alpha=1$, $a_1=0$ and $a_2=0$ we obtain
$$f=x_3^1(x_1^0x_2^0\ \oplus\ x_1^1x_2^1)\ \oplus\
x_1^0x_2^1x_3^1\ \oplus\  x_1^1x_2^0x_3^1\ =x_3^1. $$ This is the
case {\bf B.b.1}.
 \fbx
\begin{corollary}\label{c4.2}
$|G_2^3|=10$.
\end{corollary}
\Pr\ As we have noted the functions $f$ in the form (\ref{eq24}) are
symmetric with respect to  their variables. Hence there are exactly
two such functions, obtained for $\alpha=1$ and $\alpha=0$. These
functions are realized in the case {\bf B.a.2.}

Let us consider the functions  $f$ in the form (\ref{eq23}) with
$\alpha=\beta$. Then we have
$$f=x_1^0x_2^1x_3^\alpha\ \oplus\ x_1^1x_2^0x_3^\alpha\ \oplus\
x_1^\alpha x_2^\alpha x_3^0\ \oplus\  x_1^\alpha x_2^\alpha x_3^1.$$
It is easy to check that in the both cases $\alpha=1$ and $\alpha=0$
the function $f$ is symmetric. Hence there exist exactly two
functions from $P_2^2$ in the form (\ref{eq23}) with $\alpha=\beta$.
These two functions are realized in the case {\bf A.b.2.}

Finally, let us consider the functions in the form
\begin{equation}~\label{eq25}
f=x_3^\alpha(x_1^0x_2^1\ \oplus\ x_1^1x_2^0)\ \oplus\
x_1^{\neg\alpha} x_2^{\neg\alpha}.
\end{equation}
Since $f(\alpha,\beta,\neg(\alpha))=0$ and
$f(\neg(\alpha),\beta,\alpha)=1$ for all $\beta\in\{0,1\}$ it
follows that $f$ is not symmetric with respect to $x_1$ and $x_3$.
Furthermore, it is clear that $f$ is  symmetric with respect to
$x_1$ and $x_2$. Hence there are exactly six functions from $P_2^3$
in the form (\ref{eq25}). When $\alpha=1$ we obtain three function
by three permutations of the variables and the same number functions
for $\alpha=0$. These functions are realized in the cases: {\bf
A.a.2.}, {\bf B.a.1.}  and {\bf B.b.2.}
 \fbx
\begin{lemma}~~\label{l6}
Let $f=x_4^0.g(x_1,x_2,x_3)\ \oplus\ x_4^1.h(x_1,x_2,x_3)\in P_2^4$.
If $f\in G_2^4$, then  $ess(g_{i\laa j})< 2$ and $ess(h_{i\laa j})<
2$ for all $i,j,\ 1\leq i < j\leq 3$.
\end{lemma}
\Pr\  Let us suppose that the lema is false. Without loss of
generality let us assume $ess(g_{2\laa 1})\geq
 2$. If $f\in
G_2^4$, then  $x_4\notin Ess(f_{2\laa 1})$ because
$$f_{2\laa 1}=x_4^0.g_{2\laa 1}\oplus x_4^1.h_{2\laa 1}\ \mbox{ and
}\ f_{2\laa 1}(x_4=0)=g_{2\laa 1}.$$ From Theorem \ref{t2} it
follows that $g_{2\laa 1}=h_{2\laa 1}$. Let us set
$$
g:=\bigoplus_{m=0}^7a_m^{(0)}.x_1^{\alpha_1}x_2^{\alpha_2}x_3^{\alpha_3}\
\ \mbox{and   }\
 h:=\bigoplus_{m=0}^7a_m^{(1)}.x_1^{\alpha_1}x_2^{\alpha_2}x_3^{\alpha_3},
 $$
where $m=\o{\alpha_1\alpha_2\alpha_3},$ and
$$t:=a_0^{(0)}.x_1^{0}x_2^{0}x_3^{0}\ \oplus\
a_1^{(0)}.x_1^{0}x_2^{0}x_3^{1}\ \oplus\
a_6^{(0)}.x_1^{1}x_2^{1}x_3^{0}\ \oplus\
a_7^{(0)}.x_1^{1}x_2^{1}x_3^{1}. $$ Then from $g_{2\laa 1}=h_{2\laa
1}$, we obtain
$$g:=t(x_1,x_2,x_3)\oplus\big(\bigoplus_{\alpha_1\neq\alpha_2}a_m^{(0)}.
x_1^{\alpha_1}x_2^{\alpha_2}x_3^{\alpha_3}\big)\ \ \mbox{and   }$$
$$
h:=t(x_1,x_2,x_3)\oplus\big(\bigoplus_{\alpha_1\neq\alpha_2}a_m^{(1)}.
 x_1^{\alpha_1}x_2^{\alpha_2}x_3^{\alpha_3}\big).$$
 Note that
 $$f_{2\laa 1}=t_{2\laa 1}=g_{2\laa 1}=h_{2\laa 1}.$$

 If $ess(g_{2\laa 1})> 2$, then from $f_{2\laa 1}(x_4=0)=g_{2\laa
 1}$ it follows that $f\notin G_2^4$.

 Hence $ess(g_{2\laa 1})= 2$. Thus we have $\{x_1,x_3\}=Ess(g_{2\laa
 1})$. This implies
 \begin{equation}~\label{eq37}
(a_0^{(0)},a_6^{(0)})\neq (a_1^{(0)},a_7^{(0)})\ \ \mbox{and
}(a_0^{(0)},a_1^{(0)})\neq (a_6^{(0)},a_7^{(0)}).
 \end{equation}
From $x_4\in Ess(f)$ it follows that there are three numbers
$\alpha_1,\alpha_2,\alpha_3\in \{0,1\}$ such that $a_m^{(0)}\neq
a_m^{(1)}$ where $m=\o{\alpha_1\alpha_2\alpha_3}.$ Then
$\alpha_1\neq\alpha_2$. Hence we have $\alpha_1=\alpha_3$ or
$\alpha_2=\alpha_3$.

Let us assume $\alpha_1=\alpha_3$. Then the identification minor
$u=f_{3\laa 1}$ can be written as follows
$$u=a_0^{(0)}.x_1^{0}x_2^{0}\ \oplus\ a_7^{(0)}.x_1^{1}x_2^{1}\ \oplus\
x_4^0(a_2^{(0)}.x_1^{0}x_2^{1}\ \oplus\ a_5^{(0)}.x_1^{1}x_2^{0})\
\oplus\ x_4^1(a_2^{(1)}.x_1^{0}x_2^{1}\ \oplus\
a_5^{(1)}.x_1^{1}x_2^{0}).$$

Without loss of generality let us assume that $a_2^{(0)}\neq
a_2^{(1)}$, i.e. $m=\o{010}=2.$ (Alternative  opportunity is $m=5$.)
Then we have $a_2^{(0)}\neq 0$ or $a_2^{(1)}\neq 0$. Again, without
loss of generality let us assume $a_2^{(0)}=1$ and $a_2^{(1)}= 0$.
Then  $u(x_1=\alpha_1,x_2=\alpha_2)=a_2^{(0)}.x_4^0\ \oplus\
a_2^{(1)}.x_4^1$. Hence $x_4\in Ess(u)$.

On the other hand we have
$$u_1=u(x_4=0)=a_0^{(0)}.x_1^{0}x_2^{0}\ \oplus\ a_7^{(0)}.x_1^{1}x_2^{1}\ \oplus\
x_1^{0}x_2^{1}\ \oplus\ a_5^{(0)}.x_1^{1}x_2^{0}\ \ \mbox{and}$$
$$u_2=u(x_4=1)=a_0^{(0)}.x_1^{0}x_2^{0}\ \oplus\ a_7^{(0)}.x_1^{1}x_2^{1}\ \oplus\
a_5^{(1)}.x_1^{1}x_2^{0}.$$

Thus we have:\\
 If $a_0^{(0)}=a_7^{(0)}=0$ or $a_0^{(0)}=a_7^{(0)}=1$, then
$Ess(u_1)=\{x_1,x_2\}$.

Let $a_0^{(0)}\neq a_7^{(0)}$. Then  according to (\ref{eq37}) we
can assume without loss of generality that $a_0^{(0)}=1$ and
$a_7^{(0)}=0$. Now, we have:\\
If $a_5^{(0)}=1$ or $a_5^{(1)}=0$, then $Ess(u_1)=\{x_1,x_2\}$ or
$Ess(u_2)=\{x_1,x_2\}$.

Finally, if $a_0^{(0)}=1$, $a_7^{(0)}=0$, $a_5^{(0)}=0$ and
$a_5^{(1)}=1$ we have $u_1(x_1,x_2)=x_1^0$ and $u_2(x_1,x_2)=x_2^0$.

So, we have shown that $Ess(u)=\{x_1,x_2,x_4\}$. Hence $f\notin
G_2^4$, which is a contradiction.

By symmetry, we obtain the same contradiction when
$\alpha_2=\alpha_3$ and we have to use the identification minor
$v=f_{3\laa 2}$ instead of $u=f_{3\laa 1}$.
 \fbx

\begin{lemma}~~\label{l7}
Let $f=x_4^0.g(x_1,x_2,x_3)\ \oplus\ x_4^1.h(x_1,x_2,x_3)\in P_2^4$.
If $f\in G_2^4$, then $ess(g)=ess(h)=3$.
\end{lemma}
\Pr\ Let us suppose that $x_3\notin Ess(g)$ and $f\in G_2^4$.

Let $g$ and $h$ are represented as follows
$$
g:=\bigoplus_{m=0}^7a_m.x_1^{\alpha_1}x_2^{\alpha_2}x_3^{\alpha_3}\
\ \mbox{and   }\
 h:=\bigoplus_{m=0}^7b_m.x_1^{\alpha_1}x_2^{\alpha_2}x_3^{\alpha_3},
 $$
where
$m=\o{\alpha_1\alpha_2\alpha_3}=\alpha_1.2^2+\alpha_2.2+\alpha_3.$
Since $x_3\notin Ess(g)$, we obtain
\begin{equation}~\label{eq29} (a_0,a_2,a_4,a_6)=(a_1,a_3,a_5,a_7).\end{equation}

 On the other hand
$x_3\notin Ess(g)$ implies $x_3\in Ess(h)$. Hence, we have
\begin{equation}~\label{eq30}(b_0,b_2,b_4,b_6)\neq(b_1,b_3,b_5,b_7).\end{equation}

Without loss of generality let us assume that $b_0=1$ and $b_1=0$.
 Consequently,
$$u=f_{2\laa 1}= x_4^0(a_0x_1^0\ \oplus\ a_6x_1^1)\ \oplus\ x_4^1(x_1^0x_3^0\ \oplus\ b_6.x_1^1x_3^0
 \ \oplus\ b_7x_1^1x_3^1).$$
 From $u(x_1=0,x_4=1)=x_3^0$ it follows $x_3\in Ess(u)$. If $a_0=1$,
 then $u(x_1=0,x_3=1)=x_4^0$ and if $a_0=0$,
 then $u(x_1=0,x_3=0)=x_4^1$. Hence $x_4\in Ess(u)$. The proof will
 be done if we show that $x_1\in Ess(u)$. Suppose the opposite i.e. $x_1\notin Ess(u)$.
 From Theorem \ref{t2} it follows that $a_0=a_6$, $b_6=1$ and
 $b_7=0$. Then we have
$$v=f_{3\laa 1}= x_4^0[a_0.(x_1^0x_2^0\ \oplus\ x_1^1x_2^1)\ \oplus\ a_2.x_1^0x_2^1\ \oplus\ a_4.x_1^1x_2^0]\
\oplus$$ $$\oplus\
 x_4^1[x_1^0x_2^0\ \oplus\ b_2.x_1^0x_2^1
 \ \oplus\ b_5.x_1^1x_2^0].$$
 If $a_0=1$,
 then $v(x_1=1,x_2=1)=x_4^0$ and if $a_0=0$,
 then $v(x_1=0,x_2=0)=x_4^1$. Hence $x_4\in Ess(v)$. On the other
 side it is clear that $v(x_4=1):=x_1^0x_2^0\ \oplus\ b_2.x_1^0x_2^1
 \ \oplus\ b_5.x_1^1x_2^0$ is not a constant. Assume that $x_2\in
 Ess(v)$. Suppose that $x_1\notin Ess(v)$. Hence $a_0=a_2=a_4$, $b_5=1$ and
 $b_2=0$.
 Thus we obtain
$$w=f_{3\laa 2}= a_0.x_4^0\ \oplus\ x_4^1(x_1^0x_2^0\ \oplus\
b_3.x_1^0x_2^1
 \ \oplus\ b_4x_1^1x_2^0).$$

Clearly $x_4\in Ess(w)$. On the other hand it is clear that
$w(x_4=1):=x_1^0x_2^0\ \oplus\ b_3.x_1^0x_2^1
 \ \oplus\ b_4.x_1^1x_2^0$ is not a constant. Assume that $x_2\in
 Ess(w)$. Suppose that $x_1\notin Ess(w)$. Hence  $b_3=0$ and
 $b_4=1$.
  Thus finally, we obtain
 $$f= a_0.x_4^0\ \oplus\ x_4^1(x_1^0x_2^0x_3^0\ \oplus\ x_1^1x_2^0x_3^0\
 \oplus\ x_1^1x_2^0x_3^1\ \oplus\ x_1^1x_2^1x_3^0).$$
 The contradiction is $f\notin G_2^4$ because
 $f_{4\laa 2}=a_0.x_2^0\ \oplus\ x_1^1x_2^1x_3^0$.

 By analogy we conclude that $f\notin G_2^4$ for the all other cases
 generated by (\ref{eq29}) and (\ref{eq30}), which is a contradiction.
 \fbx
\begin{theorem}~~\label{t5}
Let $f\in P_2^4$. Then $f\in G_2^4$ if and only if $f=
x_4^0.g(x_1,x_2,x_3)\ \oplus\ x_4^1.h(x_1,x_2,x_3)$, with
\begin{equation}~~~\label{eq31}
g=x_3^\alpha(x_1^0x_2^0\ \oplus\ x_1^1x_2^1)\ \oplus\
x_3^{\neg(\alpha)}(x_1^0x_2^1\ \oplus\ x_1^1x_2^0),
\end{equation}
and
\begin{equation}~~~\label{eq32}
h=x_3^{\neg(\alpha)}(x_1^0x_2^0\ \oplus\ x_1^1x_2^1)\ \oplus\
x_3^{\alpha}(x_1^0x_2^1\ \oplus\ x_1^1x_2^0),
\end{equation}
for some $\alpha$, $\alpha\in\{0,1\}$.
\end{theorem}
\Pr\  $"\La"$ The proof in this direction is done in Proposition
\ref{p2}.

 $"\Ra"$ Suppose that some of the equations  (\ref{eq31}) or
 (\ref{eq32}) are not satisfied. From Lemma \ref{l6} and Lemma
 \ref{l7} there are two possible cases:

{\bf A.}
$$g=x_3^\alpha(x_1^0x_2^1\ \oplus\ x_1^1x_2^0)\
\oplus\ x_1^\beta x_2^{\beta},$$ and
$$h=x_3^{\gamma}(x_1^0x_2^0\ \oplus\ x_1^1x_2^1)\ \oplus\
x_3^{\neg(\gamma)}(x_1^0x_2^1\ \oplus\ x_1^1x_2^0).$$ Then we have
the following identification minor of $f$:
$$u=f_{4\laa 1}=x_1^0x_2^1x_3^\alpha\ \oplus\ \neg(\beta).x_1^0x_2^0\
\oplus\ x_1^1x_2^1x_3^\gamma\ \oplus\
x_1^1x_2^0x_3^{\neg(\alpha)}.$$ Since $u(x_1=0)=x_2^1x_3^\alpha\
\oplus\ \neg(\beta).x_2^0$ it follows that $\{x_2,x_3\}\subseteq
Ess(u)$. We will show that $x_1\in Ess(u)$, also.

Let $\beta=0$. If $\gamma=\alpha$, then we have
$u(x_2=0,x_3=\gamma)=x_1^0$, and if $\gamma\neq\alpha$, then we have
$u(x_2=1,x_3=\gamma)=x_1^1$.

Let $\beta=1$. If $\gamma=\alpha$, then
$u(x_2=0,x_3=\neg(\gamma))=x_1^1$, and if $\gamma\neq\alpha$, then
we have $u(x_2=1,x_3=\gamma)=x_1^1$.

Hence $x_1\in Ess(u)$ and $f\notin G_2^4$ in the case {\bf A.}

{\bf B.}
$$g=x_3^\alpha(x_1^0x_2^1\ \oplus\ x_1^1x_2^0)\
\oplus\ x_1^\beta x_2^{\beta},$$ and
$$h=x_3^\gamma(x_1^0x_2^1\ \oplus\ x_1^1x_2^0)\
\oplus\ x_1^\delta x_2^{\delta}.$$ Since $x_4\in Ess(f)$ it follows
that $g\neq h$.

Let us consider the  identification minor $u$ of $f$, also:
$$u=f_{4\laa 1}=x_1^0x_2^1x_3^\alpha\ \oplus\ \neg(\beta).x_1^0x_2^0\
\oplus\ x_1^1x_2^0x_3^\gamma\ \oplus\ \delta.x_1^1x_2^1.$$

 Since
$u(x_1=0)=x_2^1x_3^\alpha\ \oplus\ \neg(\beta).x_2^0$ it follows
that $\{x_2,x_3\}\subseteq Ess(u)$. We will prove that $x_1\in
Ess(u)$, also.

Let $\beta=\delta=0$. Then $u(x_2=1,x_3=\alpha)=x_1^0$;

Let $\beta=\delta=1$. Then $u(x_2=0,x_3=\gamma)=x_1^1$;

Let $\beta=1$ and $\delta=0$. Then $u(x_2=0,x_3=\gamma)=x_1^1$;

Let $\beta=0$ and $\delta=1$. Then
$u(x_2=1,x_3=\neg(\alpha))=x_1^1$.

Hence $x_1\in Ess(u)$ and $f\notin G_2^4$ in the case {\bf B.},
also. This is a contradiction.
 \fbx
 \begin{remark}\label{rem1}
 Note that $g$ and $h$ have to be two special functions from $G_2^3$, represented
 by the equation  (\ref{eq24}) of  Theorem \ref{t4}. Such functions can be obtained
  in the cases of the same
 theorem  {\bf B.a.2} and {\bf B.b.2}, only.
 \end{remark}
 \begin{corollary}~~\label{c5.3}
Let $f\in P_2^4$. Then $f\in G_2^4$ if and only if $f=
x_4^0.g(x_1,x_2,x_3)\ \oplus\ x_4^1.h(x_1,x_2,x_3)$, with
$$
g=x_3^\alpha(x_1^0x_2^0\ \oplus\ x_1^1x_2^1)\ \oplus\
x_3^{\neg(\alpha)}(x_1^0x_2^1\ \oplus\ x_1^1x_2^0),
$$
and $ h=\neg(g(x_1,x_2,x_3))$.
\end{corollary}
 \begin{corollary}~~\label{c5.2}
Let $f\in P_2^4$. Then $f\in G_2^4$ if and only if
$$f=
a_0.\big(\bigoplus_{\alpha_1\alpha_2\alpha_3\alpha_4\in
Od_2^4}x_1^{\alpha_1}x_2^{\alpha_2}x_3^{\alpha_3}
 x_4^{\alpha_4}\big)\oplus\ \neg(a_0).\big(
 \bigoplus_{\alpha_1\alpha_2\alpha_3\alpha_4\in Ev_2^4}x_1^{\alpha_1}x_2^{\alpha_2}x_3^{\alpha_3}
 x_4^{\alpha_4}\big). $$
\end{corollary}
\begin{corollary}\label{c5.1}
If $f\in G_2^4$ then $x_j\notin Ess(f_{i\laa j})$ for all
$i,j\in\{1,2,3,4\}\ i\neq j$.
\end{corollary}
\Pr\ The three corollaries, above can be proved by immediate
checking of the both functions from  $G_2^4$, obtained in Theorem
\ref{t5}. \fbx

\begin{theorem}\label{t6}
A Boolean function $f\in P_2^n$, depending on $n$ essential
variables with $n\geq 4$, has essential arity gap 2 if and only if
\begin{equation*}~\label{eq1} f= \bigoplus_{\alpha_1\ldots\alpha_n\in
Od_2^n}x_1^{\alpha_1}\ldots
 x_n^{\alpha_n}\quad \mbox{ or }\quad
 f= \bigoplus_{\alpha_1\ldots\alpha_n\in Ev_2^n}x_1^{\alpha_1}\ldots
 x_n^{\alpha_n}.\end{equation*}
\end{theorem}
\Pr\ $"\La"$ In this direction the proof is done by Proposition
\ref{p2}.

$"\Ra"$
 We will proceed by induction on $n$. If $n=4$ the theorem is
true because of Theorem \ref{t5}. Suppose that if $4\leq n\leq l$
and $f\in G_2^n$, then
$$f= \bigoplus_{\alpha_1\ldots\alpha_n\in
Od_2^n}x_1^{\alpha_1}\ldots
 x_n^{\alpha_n}\quad \mbox{ or }\quad
 f= \bigoplus_{\alpha_1\ldots\alpha_n\in Ev_2^n}x_1^{\alpha_1}\ldots x_n^{\alpha_n}.$$
 Let $f\in G_2^{l+1}$. Hence $f$ can be presented as follows
 $$f=x_{l+1}^0.g(x_1,\ldots,x_l)\oplus x_{l+1}^1.h(x_1,\ldots,x_l).$$
 In the same way as in Lemma \ref{l6} and Lemma \ref{l7} it
 can be proved  that $g,h\in G_2^{l}$. By the inductive
 supposition $g$ and $h$ are  functions of the forms
$$ \bigoplus_{\gamma_1\ldots\gamma_l\in
Od_2^l}x_1^{\gamma_1}\ldots
 x_l^{\gamma_l}\quad \mbox{ or }\quad
 \bigoplus_{\gamma_1\ldots\gamma_l\in
Ev_2^l}x_1^{\gamma_1}\ldots
 x_l^{\gamma_l},$$
 with $g\neq h$.
Note that  $g$ and $h$ are not constants because  $ess(f)=n\geq 4$.
Hence $Ess(g_{i\laa j})=Ess(h_{i\laa j})$ for $i,j\in\{1,\ldots,l\}$
and $ i\neq j$. Assume that
$$g= \bigoplus_{\gamma_1\ldots\gamma_l\in
Od_2^l}x_1^{\gamma_1}\ldots
 x_l^{\gamma_l}\quad \mbox{ and }\quad
 h= \bigoplus_{\delta_1\ldots\delta_l\in Ev_2^l}x_1^{\delta_1}\ldots x_l^{\delta_l}. $$
Consequently
$$f=x_{l+1}^0.(\bigoplus_{\gamma_1\ldots\gamma_l\in
Od_2^l}x_1^{\gamma_1}\ldots
 x_l^{\gamma_l})\oplus
 x_{l+1}^1.(\bigoplus_{\delta_1\ldots\delta_l\in Ev_2^l}x_1^{\delta_1}\ldots x_l^{\delta_l})=$$
$$= \bigoplus_{\alpha_1\ldots\alpha_{l+1}\in
Od_2^{l+1}}x_1^{\alpha_1}\ldots
 x_{l+1}^{\alpha_{l+1}}.$$
 The case $g=h$ is impossible because  $ess(f)=l+1$, but  the
 replacement of $g$ and $h$ will produce the function
 $$f= \bigoplus_{\alpha_1\ldots\alpha_{l}\in
Ev_2^{l}}x_1^{\alpha_1}\ldots
 x_{l}^{\alpha_{l}},$$ which does not depend on $x_{l+1}$.
 \fbx
\begin{corollary}\label{c6.3}
A Boolean function $f\in P_2^n$, which essentially depends on $n$
variables with $n> 4$, has essential arity gap 2 if and only if
$$f= x_n^0.g(x_1,\ldots,x_{i},\ldots,x_{n-1})
\oplus
x_n^1.g(x_1,\ldots,x_{i-1},\neg(x_{i}),x_{i+1},\ldots,x_{n-1}),$$
where $g\in G_2^{n-1}$ and $i\in\{1,\ldots,n-1\}$.
\end{corollary}
\Pr\ If
$$g= \bigoplus_{\gamma_1\ldots\gamma_{n-1}\in
Od_2^{n-1}}x_1^{\gamma_1}\ldots
 x_{n-1}^{\gamma_{n-1}}\quad \mbox{ and }\quad
 h= \bigoplus_{\gamma_1\ldots\gamma_{n-1}\in
Od_2^{n-1}}x_1^{\gamma_1}\ldots
 x_{n-1}^{\gamma_{n-1}},$$
 then $\neg(g)=h$ and $\neg(h)=g$ for all $l\geq 4$. On the other
 hand, for each $i\in\{1,\ldots,n-1\}$, we have
  $$\neg(g)= \bigoplus_{\gamma_1\ldots\gamma_{n-1}\in
Od_2^{n-1}}x_1^{\gamma_1}\ldots
x_{i-1}^{\gamma_{i-1}}\neg(x_{i}^{\gamma_{i}})x_{i+1}^{\gamma_{i+1}}\ldots
 x_{n-1}^{\gamma_{n-1}}.$$
\fbx
\begin{corollary}\label{c6.2}
$|G_2^n|=2$ for each $n, n\geq 4$.
\end{corollary}

One of the most important problems concerning the essential arity
gap is to calculate the number of all functions from $P_2^n$, which
depend essentially  on at most $n$ variables and which have the
maximum gap i.e. with gap equal to $2$. The next theorem gives
answer of that problem. It summarize the results obtained above in
the paper.

Let us denote by $H_n$ the set of all functions in $P_2^n$, which
have gap equal to $2$ i.e.
$$H_n:=\bigcup_{m=2}^nG_2^m\ \ \mbox{and}\ \ h_n:=|H_n|.$$

\begin{theorem}\label{t12} The following combinatorial equations are
held:

 $(i)$ $h_2=6$;

 $(ii)$ $h_3=28$;

 $(iii)$ $h_n=3.{n\choose 2}+5.{n\choose 3}+2^{n+1}-2n-2$, when $n\geq 4$;

\end{theorem}
\Pr\

$(i)$\ \  follows from Corollary \ref{c3.1} of Theorem \ref{t3};

 $(ii)$ Let $X_3=\{x_1,x_2,x_3\}$. There are
 $6.{3\choose 2}$ Boolean functions with essential arity gap equal to
 2,
 which depend essentially on $2$ variables from $X_3$, according to
 Corollary \ref{c3.1} of Theorem \ref{t3}.

 From  Corollary
\ref{c4.2} of Theorem \ref{t4} it follows that there are $10$
Boolean functions with essential arity gap equal to 2,
 which depend essentially on all $3$ variables from $X_3$. Hence
 $h_3=6.3+10=28$.

$(iii)$  Let $X_n=\{x_1,\ldots,x_n\}$, $n\geq 4$. There are
 $6.{n\choose 2}$ Boolean functions with essential arity gap equal to
 2,
 which depend essentially on $2$ variables from $X_n$, according to
 Corollary \ref{c3.1} of Theorem \ref{t3}.

There are
 $10.{n\choose 3}$ Boolean functions with essential arity gap equal to
 2,
 which depend essentially on $3$ variables from $X_n$, according to
 Corollary
\ref{c4.2} of Theorem \ref{t4}.

Finally, for each $m,$ $3<m\leq n$ there are $2.{n\choose m}$
Boolean functions with essential arity gap equal to 2,
 which depend essentially on $m$ variables from $X_n$, according to
Corollary \ref{c6.2} of Theorem \ref{t6}.

Hence we have

$$h_n= 6.{n\choose 2}+10.{n\choose 3}+ 2.\big[{n\choose 4}+{n\choose 5}+\ldots+{n\choose n}\big]=$$
$$=3.{n\choose 2}+5.{n\choose 3}+2^{n+1}-2n-2.$$

 \fbx

\end{document}